\documentclass[FBSmath,ecsub]{FBSsuppl}
\usepackage{epsfig}
\usepackage{amsmath}
\usepackage{amsfonts}
\usepackage{amssymb}
\usepackage{color}

\definecolor{Conf}{named}{Black}
\definecolor{GBE}{named}{Black}
\definecolor{OGE}{named}{Black}
\definecolor{Text-normal}{named}{Black}

\title{Baryons as Relativistic Three-Quark Systems}

\author{W. Plessas}

\institute{Institut f\"ur Theoretische Physik,
Universit\"at Graz, Universit\"atsplatz 5,
A-8010 Graz, Austria}

\begin{document}
\maketitle
\begin{abstract}
We discuss some recent developments in the description of baryons
as three-quark systems within relativistic constituent quark models.
In particular we address the issues of excitation spectra,
electroweak structure, and mesonic resonance decays. The necessities
of implementing simultaneously the symmetries of low-energy quantum
chromodynamics and of relativistic invariance are emphasized.
\end{abstract}
\section{Introductory Remarks}

The consideration of few-quark systems has become a very exciting 
topic in few-body physics. On the one hand it represents a big 
challenge for modern few-body methods to solve few-quark problems, 
since the corresponding bound and continuum states have to be
solved in a consistent 
relativistic framework. Thus, in order to treat hadron phenomena on a 
microscopic level the formalism of calculation has to be essentially 
relativistic. On the other hand the issues of quark dynamics are very 
important. In principle, the microscopic description of hadronic 
bound states and reactions amounts to solving quantum
chromodynamics (QCD) at low energies. Evidently, this is not strictly
possible nowadays, and one must resort to effective theories and/or 
models. For any of such approaches, however, it is required to 
implement the essential properties of low-energy QCD, which are still 
not known in a satisfactory manner. In working with few-quark problems 
it is an intriguing task to find out the forces binding the 
constituents of hadrons and finally governing their behaviours.
Evidently it is of eminent interest from many aspects of nuclear and 
subnuclear physics to be able to describe and to understand the 
properties (and reactions) of hadrons in a microscopic manner
on the basis of the 
fundamental theory of strong (and also electroweak) interaction.

A very powerful tool for low- and intermediate-energy hadron physics 
consists in employing constituent quark models (CQMs). They have 
already witnessed a long and reasonably successful development. 
Starting out from the naive nonrelativistic quark model of hadrons in 
the 60's one has constantly improved the quark-quark dynamics and one 
has learned how to apply a relativistic formalism\footnote{By 
`relativistic' we mean that a model should at least meet the 
requirements of Poincar\'e invariance. Thus it need not be 
field-theoretic, with infinitely many degrees of freedom, but can also 
be based on a covariant mass operator or Hamiltonian, defined on a 
Hilbert space with a finite number of degrees of freedom.}.
At present we have 
several refined CQMs available, which at least prove the approach to
be promising. Of course, they advocate various types of dynamics
and use different calculational methods, what leads to partly 
dissimilar results. However, this just makes up the flavour of present
research in this field. By considering more and more hadron 
processes within the concept of CQMs one will gradually find out which 
is the most promising variant of the existing models.

In hadronic physics one would like to have an effective theory/model 
of QCD that is able to deal with all phenomena in a certain energy 
domain (at low and intermediate energies, say). It is not sufficient 
to consider just a limited aspect, for instance, only spectroscopy. 
This is especially true also for CQMs. Therefore, any realistic CQM 
should be subject to as many tests as possible. A model that has 
proven to be successful in describing the hadron spectra (bound state 
and resonance energies) should then be applied to hadronic reactions 
in order to check its global usability (notably also the wave 
function behaviour).

In this context, we shall here mainly discuss results of the 
relativistic CQM whose 
dynamics is based on Goldstone-boson exchange (GBE) vis-\`a-vis 
corresponding results from other types of relativistic CQMs. In 
particular, we shall consider the GBE CQM by the Graz group
\cite{Glozman:1998ag,Glozman:1998fs} and compare its results to the 
ones by a typical representative of one-gluon-exchange (OGE) CQMs, 
namely the relativistic variant of the Bhaduri-Cohler-Nogami
(BCN) CQM \cite{Bhaduri:1981pn} as parametrized in ref.
\cite{Theussl:2000s}, and by the CQM relying on instanton-induced
(II) forces \cite{Loring:2001kx}. We shall address the aspects of 
light and strange baryon spectroscopy, nucleon electroweak structure, 
electric radii and magnetic moments of octet and decuplet baryon 
ground states, and pionic decays of nucleon and $\Delta$ resonances.

\section{Baryon Spectroscopy}

\begin{figure}[ht]
\epsfig{file=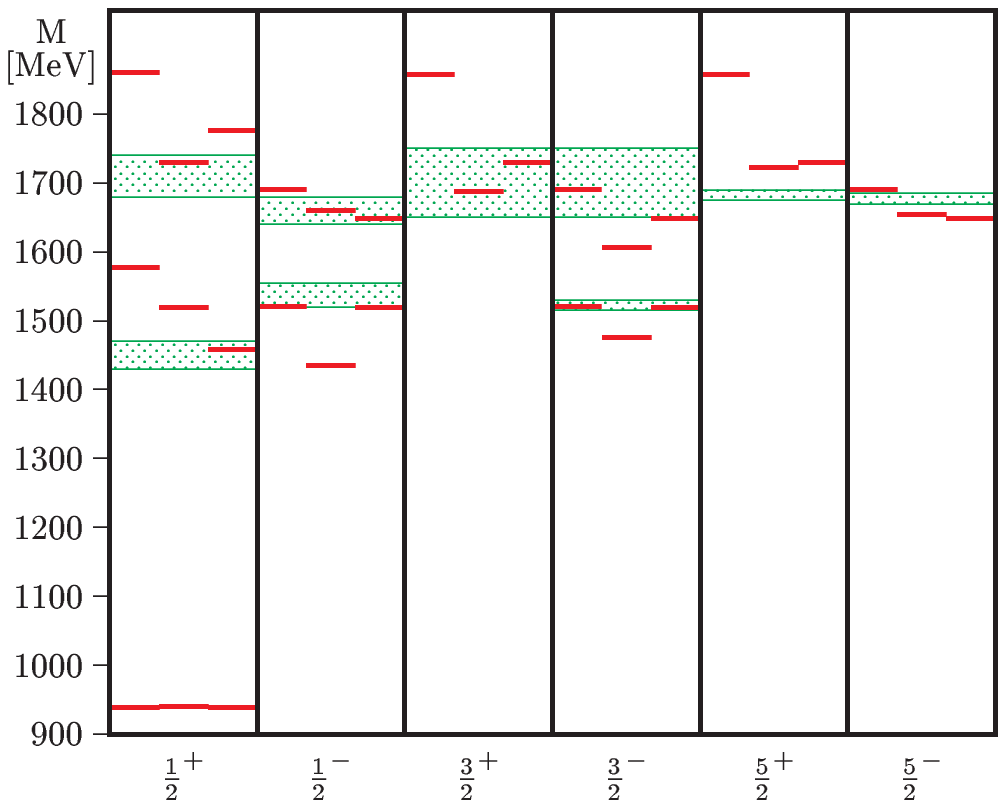,width=10cm}
\caption[]{Nucleon excitation spectra of three different types of 
relativistic CQMs. In each column the left horizontal lines represent 
the results of the BCN OGE CQM \protect\cite{Theussl:2000s}, the
middle ones of the II CQM (Version A) \protect\cite{Loring:2001kx},
and the right ones of the GBE CQM
\protect\cite{Glozman:1998ag}. The shadowed boxes
give the experimental data with their uncertainties after the
latest compilation of the PDG \protect\cite{PDG2002}.}
\vspace{-2mm}
\end{figure}
\begin{figure}[ht]
\epsfig{file=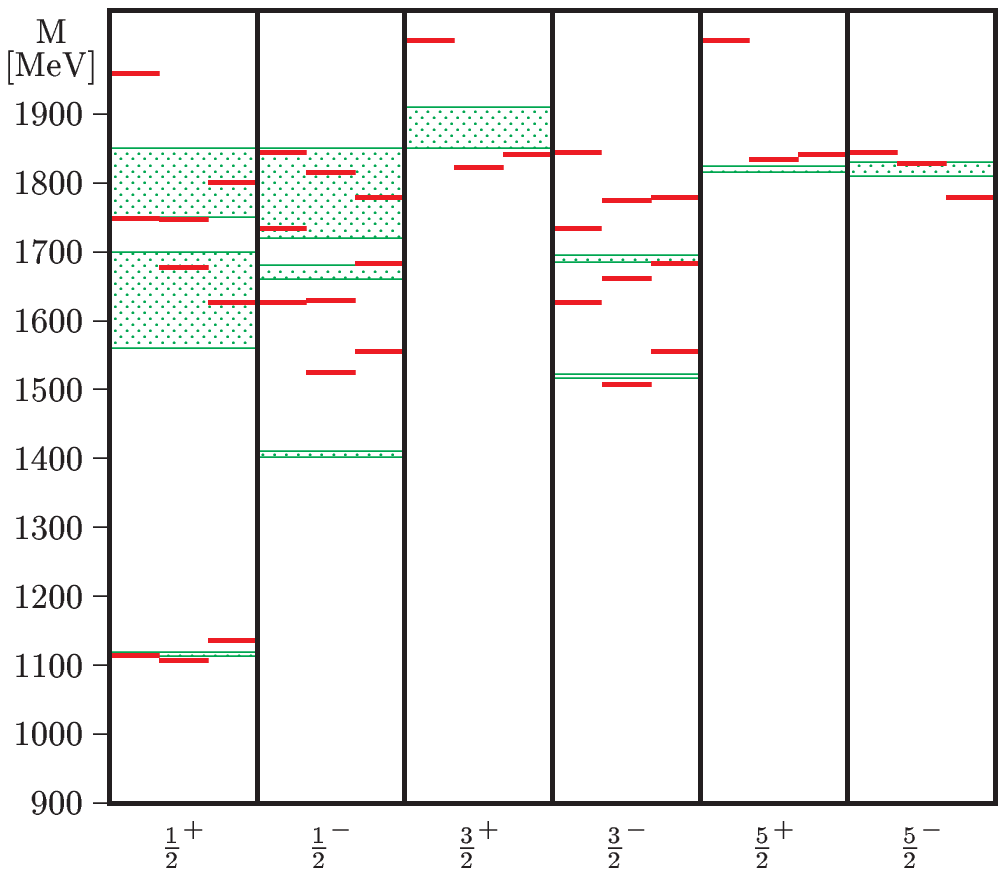,width=10cm}
\caption[]{$\Lambda$ excitation spectra. Same caption as in Fig. 1.}
\vspace{-2mm}
\end{figure}

The baryon spectra as provided by CQMs have been discussed from many 
viewpoints, especially also because of the big interest from 
the experimental side. While the description was rather qualitative in 
the early stages of CQMs, it is clear that considerable improvements 
have been achieved in the past years. Finally one has met the demand 
of describing all the light- and strange baryon spectra within one 
and the same framework. Due to the very nature of these spectra it
is immediately evident that one needs flavor-dependent forces. As a 
result, the OGE CQM being flavor-independent
encounters difficulties in describing simultaneously the $N$ and
$\Lambda$ spectra, 
in addition to the notorious shortcomings in the level orderings of 
positive- and negative-parity states in the $N$ (and tentatively 
$\Delta$) spectrum. The reasons have already been discussed in much 
detail in the literature (see, e.g., ref. \cite{Glozman:1998fs}). Here 
we only recall the typical behaviour through a comparison of the 
OGE, II, and GBE CQMs for the $N$ and $\Lambda$ spectra in Figs. 1 and 
2 (the complete spectra can be found in the original refs.
\cite{Theussl:2000s}, \cite{Loring:2001kx}, and 
\cite{Glozman:1998ag}, respectively). This comparison actually
shows the (most striking) current problems still prevailing in
the various types of CQMs. Notably, all of them encounter severe 
shortcomings in the $\Lambda$ spectrum, especially with regard to 
the $\Lambda$(1405), the first excitation above the ground state. 
Furthermore the OGE and the II CQMs are left with the wrong level 
orderings of the first positive- and negative-parity excitations above 
the nucleon ground state. Only the GBE CQM can reproduce both the 
$N^{*}$(1440) ${\frac{1}{2}}^{+}$ and the $N^{*}$(1535)
${\frac{1}{2}}^{-}$ states in the right places, in accordance with
experiment. Except for the open problem with $\Lambda$(1405), it thus 
represents the only CQM that is able to describe in a unified 
framework all light and strange baryon spectra in reasonable agreement 
with experiment.

With regard to the problem of the correct level ordering of positive- 
and negative-parity states the following remark is in order. It has 
repeatedly been claimed above all by Valcarce et al. that a proper
description, especially of the $N$ and $\Delta$ spectra, 
can also be obtained within a hybrid (i.e. one-gluon exchange plus 
meson exchange) CQM keeping a predominant contribution from the OGE 
interaction. In the first paper of the Salamanca-Valencia
group \cite{VGFV},
where such a claim was originally made, they used a variant of their 
CQM with a value $\alpha_{s}=0.485$ of the effective quark-gluon 
coupling and a cut-off parameter $r_{0}=0.0367$ fm in the OGE
part of the hyperfine interaction. Due to a 
nonconverged variational calculation of the nucleon spectrum they 
obtained a wrong level structure, with the positive-parity 
$N^{*}(1440)$ resonance practically degenerate with the 
negative-parity $N^{*}(1535)$ state. Upon recalculating this model in 
a converged variational calculation and also in a rigorous Faddeev 
approach, the Graz group established the right $N$ and $\Delta$ 
spectra and found them in disagreement with the original paper
\cite{VGFV} and thus also at variance with phenomenology
\cite{Glozman:1998fs}. The problem 
was reiterated in ref. \cite{GPPVW99}, where also some technical 
details about treating the $\delta$-function smearing of the 
color-magnetic interaction in three-quark calculations were 
elucidated. Still, in ref. \cite{FGV} a similar claim as before was 
made by the Salamanca-Valencia group, then with a different
parametrization of their CQM, namely with the parameters
$\alpha_{s}=0.65$ and $r_{0}=0.8$ fm in the OGE Part. As it turned 
out in another recalculation by the Graz group, the results reported 
in ref. \cite{FGV} were again not converged. Similar errors were 
repeated in a subsequent paper by Garcilazo et al. \cite{GVF01a}.
There, some further modified versions of the Salamanca-Valencia
CQM were considered reporting 
results for the $N$ and $\Delta$ spectra calculated in the framework 
of Faddeev equations. One claimed a reasonable reproduction of the 
ground-state and resonance levels, in particular of the Roper 
$N^{*}(1440)$ state, while maintaining a considerable contribution from
the OGE part in the $Q$-$Q$ hyperfine interaction. Upon checking these 
results the Graz group found another time that they were not
satisfactorily accurate. This was established by applying on the
one hand the stochastic variational method (SVM) \cite{svm}
and on the other hand a modified 
type of Faddeev approach as in ref. \cite{papp}; the latter was 
especially designed to treat infinitely rising confining potentials 
along with Faddeev equations and leads to improved convergence.
Only in another subsequent paper \cite{GVF01b} Garcilazo et al.
produced practically correct $N$ and $\Delta$ spectra by including
into their Faddeev calculation enough angular momentum states.
As is clearly seen from the figures given in that work, a proper 
ordering of the positive- and negative-parity levels is not achieved 
with neither one of the different versions of the Salamanca-Valencia
hybrid CQM. On the contrary, the negative-parity $N^{*}(1535)$ falls 
clearly below the positive-parity $N^{*}(1440)$ Roper resonance, 
leaving the $N$ spectrum (and in addition also the $\Delta$ spectrum) 
at variance with experiment.

Now, there has also been a contribution to this conference by A. 
Valcarce. Therein it is stated again that `a quite reasonable 
description of the baryon spectrum' is obtained with the 
Salamanca-Valencia CQM. By inspecting fig. 5 of this contribution 
\cite{Valcarce} (which shows basically correct results but 
due to missing higher states rather incomplete excitation
spectra) one cannot find this claim confirmed by any means. 
The negative-parity ${\frac{1}{2}}^{-}$ resonances in both the
$N$ and $\Delta$ spectra lie far below the experimental values and 
notably also below the respective first positive-parity excitations 
above the ground states. In addition, the spectrum shown in fig. 5 
of ref. \cite{Valcarce} does not belong to the parameter set quoted
in table 1 of the same paper. Rather it belongs to the parameter set 
given in table 3 of ref. \cite{GVF01b}, which produces even inferior
spectra with regard to the relative positions of positive- and 
negative-parity excitations (cf. figs. 1 and 2 in ref. 
\cite{GVF01b}); in addition the first ${\frac{3}{2}}^{+}$ and
${\frac{5}{2}}^{+}$ $N^{*}$ states (not shown neither in ref. \cite{GVF01b}
nor by Valcarce in the contribution to this conference) both lie more 
than 100 MeV below the experimental levels! Thus the argumentation
in Valcarce's contribution is badly misleading in connection with the
performance of the Salamanca-Valencia CQM in baryon spectroscopy. The 
truth is that correct $N$ and $\Delta$ spectra have so far not been 
achieved in neither one of the presented parametrizations of this
particular CQM. In principle, within a pure OGE CQM or a
hybrid model that relies on a predominant
contribution from OGE in the hyperfine interaction it is not possible
to describe specifically the first positive- and negative-parity 
excitations in the $N$ and $\Delta$ spectra, the $N$-$\Delta$ 
splitting, and the $\Lambda$ spectrum correctly at the same time.
This fact can be readily derived from the spin-isospin symmetry of
the color-magnetic interaction \cite{Glozman:1998fs,glozrisk,gloz}.

Obviously a correct quantitative description of all light and strange
baryon spectra in a unified relativistic framework is of importance
for applications of the CQM wave functions in dynamical processes 
involving ground states and resonances. Certain constraints are,
of course, imposed by the spectra on the spin-flavor symmetry of
the $Q$-$Q$ interaction, and the various CQMs may not yet be
complete in these respects\footnote{See, e.g., the attempts of
extending the GBE CQM to include further force components
\protect\cite{GKPSW:2002}.}. However, one must 
also bear in mind that the description of resonances in terms of 
\{$QQQ$\} bound states (without explicit couplings to decay channels) 
might be totally inadequate (see Sec. 4 below).

\section{Electroweak Nucleon/Baryon Structure}

The particular relativistic CQMs considered in the previous
section have already been tested for nucleon form factors and, partly,
also for electric radii and magnetic moments of the light and strange
baryon ground states
\cite{Wagenbrunn:2000es,WBGKPR:2002,BPW:2002a,BPW:2002b}.
For the cases of the OGE and 
GBE CQMs one obtained manifestly covariant results along the point-form 
spectator approximation (PFSA) whereas the predictions of the II CQM
were obtained also with one-body currents but in a Bethe-Salpeter
approach \cite {Merten:2002nz}. In fig. 3 we present a 
comparison of the results for the elastic proton and neutron 
electroweak form factors. The electric radii and magnetic moments of 
the nucleons, and likewise of all other measured octet and decuplet
ground states, are quoted in tables 1 and 2.
\begin{figure}
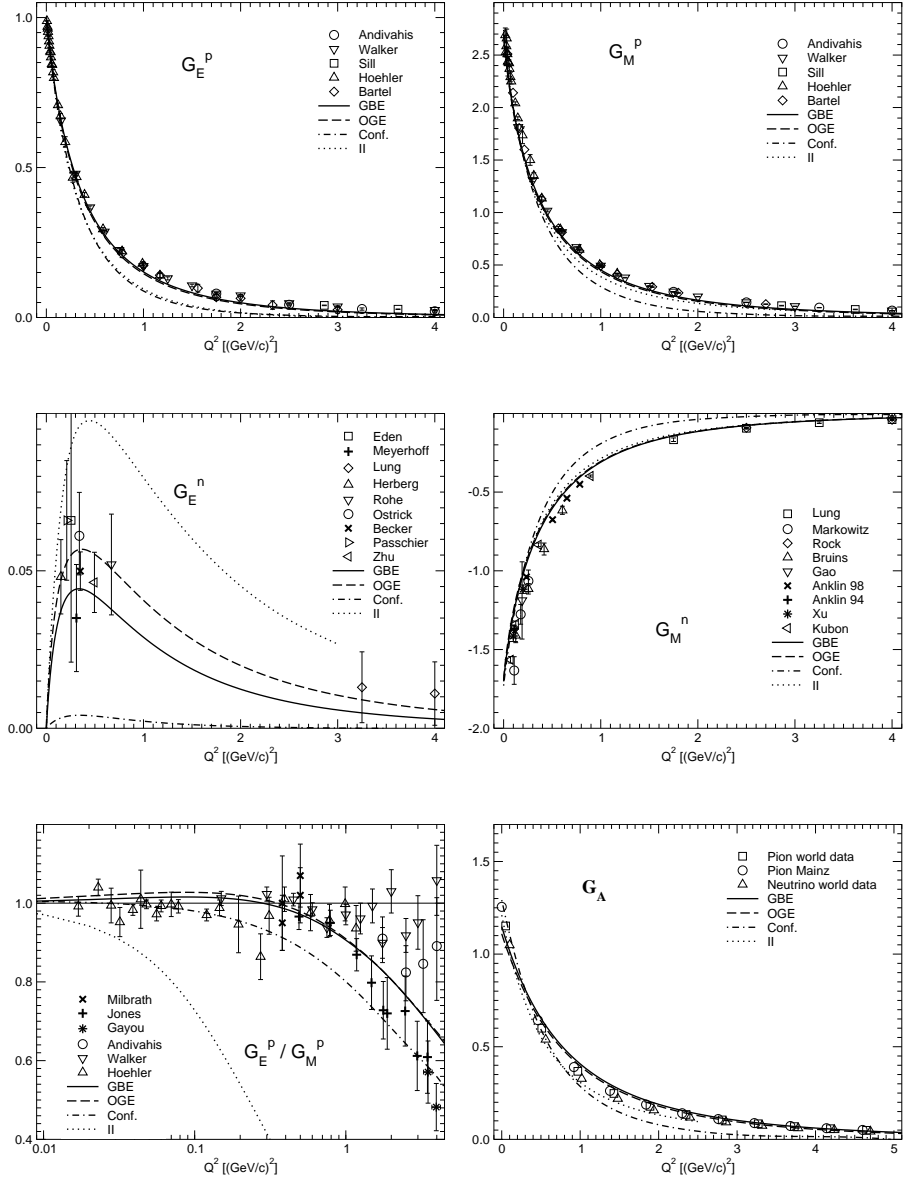

\parbox{0.5\hsize}{
\begin{center}
\includegraphics[width=1.02\hsize,bb=35 14 339 253,clip=]{gep_lisbon.eps}
\vspace{2mm}

\includegraphics[width=1.02\hsize,bb=35 14 339 253,clip=]{gen_lisbon.eps}
\vspace{2mm}

\includegraphics[width=1.02\hsize,bb=35 14 339 253,clip=]{gep_gmp_lisbon.eps}
\end{center}}
\parbox{0.5\hsize}{
\begin{center}
\includegraphics[width=1.02\hsize,bb=35 14 339 253,clip=]{gmp_lisbon.eps}
\vspace{2mm}

\includegraphics[width=1.02\hsize,bb=35 14 339 253,clip=]{gmn_lisbon.eps}
\vspace{2mm}

\includegraphics[width=1.02\hsize,bb=35 14 339 253,clip=]{ga_lisbon.eps}
\end{center}}
\vspace*{-0.2cm}
\caption{Predictions of different CQMs for the nucleon electromagnetic and
axial form factors. The solid and dashed lines represent the PFSA results
for the GBE CQM and the BCN OGE CQM, respectively; the dash-dotted lines
refer to the case with confinement only. The dotted lines show the
results of the II CQM within the Bethe-Salpeter approach after
ref. \cite{Merten:2002nz}.}
\end{figure}
\begin{table}[hb]
\label{tab1}
\caption{Predictions for electric radii of the proton, neutron, and the
$\Sigma^-$ baryon as obtained in PFSA from the GBE CQM~\cite{Glozman:1998ag}
in comparison to the case with the confinement potential only and
results from the BCN OGE CQM~\cite{Theussl:2000s}.}
\begin{tabular}[b]{|c|ccc|c|}
\hline
Baryon & \multicolumn{3}{c|}{$r_E^2$ [fm$^2$]} & 
\multicolumn{1}{c|}{Experiment}\\
& \multicolumn{1}{c}{\mbox{\color{GBE}GBE CQM}} & 
\multicolumn{1}{c}{\mbox{\color{Conf}CONF}} & \multicolumn{1}{p{2cm}|}
{\mbox{\color{OGE}OGE CQM}} & \cite{PDG2002} \\[4pt]
\hline
$p$ & ${\color{GBE}0.8176}$ & ${\color{Conf}0.7595}$ & ${\color{OGE}0.8859}$ &
$0.757\pm 0.014$\\
$n$ & ${\color{GBE}-0.1332}$ & ${\color{Conf}-0.0088}$ & ${\color{OGE}-0.1958}$ &
$-0.1161\pm 0.0022$\\[4pt]
\hline
$\Sigma^{-}$ & ${\color{GBE}0.4946}$ & ${\color{Conf}0.5329}$ & 
${\color{OGE}0.4440}$ & $0.61\pm 0.15$\\
\hline
\end{tabular}
\end{table}

In general, the relativistic results appear quite reasonable in
comparison to experiments, regarding all observables considered here.
This is the more remarkable since in all instances only one-body
operators were employed for the electromagnetic and axial currents.
Of course, one must use realistic nucleon wave functions, 
notably with mixed-symmetry spatial components included, in order to 
describe especially also the neutron and similar ground
states\footnote{With a spatially
symmetric wave function, such as the one from a confinement potential
alone, one would fail in reproducing, in particular, the neutron
electric form factor, cf. the dash-dotted line in left middle pane of 
fig. 3, which produces essentially no structure.}.
Beyond that the specific dynamical ingredients in the wave functions
are of secondary importance while relativistic (boost) effects
turn out to be very important. This is true for the results in 
point-form quantum mechanics as well as for the ones from the 
Bethe-Salpeter approach. Especially the GBE and OGE CQMs produce quite 
similar results in PFSA. But also the predictions of the II CQM are 
qualitatively similar even though they were calculated in a 
completely different relativistic formulation. Thus, it has become
quite evident that kinematical relativistic effects in any case
play a big role, and every nonrelativistic theory for electroweak
form factors must be considered as inadequate
\cite{Wagenbrunn:2000es,BPW:2002a,Merten:2002nz}.
\begin{table}[ht]
\label{tab2}
\caption{Predictions for magnetic moments of the octet baryons as well
as $\Delta^+$, $\Delta^{++}$, and $\Omega^-$ as obtained in PFSA
from the GBE CQM~\cite{Glozman:1998ag} in comparison to the
case with the confinement potential only and results from the
BCN OGE CQM~\cite{Theussl:2000s}.}
\begin{tabular}[h]{|c|ccc|c|}
\hline 
Baryon & \multicolumn{3}{c|}{$\mu$ [n.m.]} & \multicolumn{1}{c|}{Experiment}\\
& \multicolumn{1}{c}{\mbox{\color{GBE}GBE CQM}} & 
\multicolumn{1}{c}{\mbox{\color{Conf}CONF}} & \multicolumn{1}{p{2cm}|}
{\mbox{\color{OGE}OGE CQM}} & \multicolumn{1}{c|}{\cite{PDG2002},
\cite{Kotulla:02}}\\[6pt]
\hline
$p$ & ${\color{GBE}2.6980}$ & ${\color{Conf}2.6478}$ & ${\color{OGE}2.7257}$ &
2.793\\
$n$ & ${\color{GBE}-1.7004}$ & ${\color{Conf}-1.7287}$ & ${\color{OGE}-1.6951}$
& -1.913\\
\hline
$\Lambda$ & ${\color{GBE}-0.5894}$ & ${\color{Conf}-0.5802}$ & 
${\color{OGE}-0.5883}$ & -0.613\\
\hline
$\Sigma^{0}$ & ${\color{GBE}0.7001}$ & ${\color{Conf}0.6865}$ & 
${\color{OGE}0.6571}$ & -\\
$\Sigma^{+}$ & ${\color{GBE}2.3372}$ & ${\color{Conf}2.2294}$ & 
${\color{OGE}2.2046}$ & 2.458\\
$\Sigma^{-}$ & ${\color{GBE}-0.9371}$ & ${\color{Conf}-0.8564}$ & 
${\color{OGE}-0.8905}$ & -1.160\\
\hline
$\Xi^{0}$ & ${\color{GBE}-1.2744}$ & ${\color{Conf}-1.2927}$ & 
${\color{OGE}-1.2717}$ & -1.250\\
$\Xi^{-}$ & ${\color{GBE}-0.6663}$ & ${\color{Conf}-0.5490}$ & 
${\color{OGE}-0.5657}$ & -0.6507\\
\hline
$\Delta^{+}$ & ${\color{GBE}2.0827}$ & ${\color{Conf}2.0880}$ & 
${\color{OGE}2.0708}$ & $2.7\pm 1.5\pm 1.3$\\
$\Delta^{++}$ & ${\color{GBE}4.1654}$ & ${\color{Conf}4.1761}$ & 
${\color{OGE}4.1417}$ & $3.7-7.5$\\
\hline
$\Omega^-$ & ${\color{GBE}-1.5907}$ & ${\color{Conf}-1.5939}$ & 
${\color{OGE}-1.5762}$ & -2.0200\\
\hline
\end{tabular}
\end{table}

While the final insight has certainly not yet been gained, in view of 
the conceptual incompleteness of the calculations managed so far, the 
description of the nucleon (ground state) properties within CQMs 
nevertheless appears promising. Still, a number of further
investigations, e.g., employing two- and three-body currents within a 
relativistic framework, will be necessary in order to further manifest
the theoretical understanding of the nucleon structure at low momentum
transfers.

\section{Strong Decays of Baryon Resonances}

Once having considered dynamical reactions involving the ground state(s),
it is natural to proceed to the baryon resonances in order to further
test the CQMs. Despite a number of investigations available in the 
literature, unfortunately, not many evidences that can really be
considered reliable have so far been gained. In view of the experiences
made with nucleon form factors and the enormous relativistic effects
observed there, one must take any nonrelativistic study of resonances
with considerable doubt. In this connection it is certainly not
surprising that, notably, the mesonic baryon resonance decays are by
far not yet understood. Theoretical predictions obtained hitherto with
genuine CQM wave functions (i.e. without introducing additional 
phenomenological ingredients beyond the CQMs) compare quite
unfavourably with existing experiments (see, e.g., ref.
\cite{Theussl:2000s}).
\renewcommand{\arraystretch}{1.5}
\begin{table}[ht]
\caption{Predictions for pionic decay widths by the GBE 
CQM~\cite{Glozman:1998ag} along the EEM in PFSA in comparison to 
experiment, an analogous calculation with the BCN OGE
CQM~\cite{Theussl:2000s}, and results from a nonrelativistic
EEM approach.}
\label{tab3}
\begin{tabular}{|c|r|c|c|c|c|}
\hline
       &      &  \multicolumn{2}{|c|}{Rel. PFSA}
       &  \multicolumn{2}{|c|}{Nonrel. EEM } \\
Decays & Experiment~\cite{PDG2002} & \multicolumn{2}{|c|}{\small CQM} &
\multicolumn{2}{|c|}{\small GBE CQM} \\
       &  & {\small GBE} & {\small OGE} & {\small dir} & {\small dir+rec}  \\
\hline
{\small $N^{\star}_{1440}\rightarrow \pi N_{939} $}
	&  $\left(227\pm 18\right)_{-59}^{+70}$
	         &  $30.3$
             &  $37.1$
			 &  $4.85$
			 &  $6.16$ \\
\hline
{\small $N^{\star}_{1520}\rightarrow \pi N_{939} $}
	 & { $\left(66\pm 6\right)_{-5}^{+9}$  }
             &  $16.9$
			 &  $16.2$
			 &  $22.0$
			 &  $38.3$
\\
\hline
{\small $N^{\star}_{1535}\rightarrow \pi N_{939} $ }
	 & { $ \left(67\pm 15\right)_{-17}^{+55}$ }
             &  $93.2$
			 &  $122.8$
			 &  $24.3$
			 &  $574.3$
\\
\hline
{\small $N^{\star}_{1650}\rightarrow \pi N_{939}$ }
	 & { $ \left(109\pm 26 \right)_{-3}^{+36}$ }
             &  $28.8$
			 &  $38.3$
			 &  $11.3$
			 &  $160.3$
\\
\hline
{\small $N^{\star}_{1675}\rightarrow \pi N_{939} $}
	 & { $ \left(68\pm 8\right)_{-4}^{+14}$ }
             &  $5.98$
			 &  $6.20$
			 &  $7.65$
			 &  $15.1$
\\
\hline
{\small $N^{\star}_{1700}\rightarrow \pi N_{939} $ }
	 & { $ \left(10\pm 5\right)_{-3}^{+3}$ }
             &  $0.91$
			 &  $1.19$
			 &  $1.43$
			 &  $2.87$
\\
\hline
{\small $N^{\star}_{1710}\rightarrow \pi N_{939} $}
	 & { $\left(15\pm 5\right)_{-5}^{+30}$  }
             &  $4.06$
			 &  $2.28$
			 &  $23.4$
			 &  $5.95$
\\
\hline
{\small $\Delta_{1232}\rightarrow \pi N_{939} $}
	 & { $\left(119\pm 1 \right)_{-5}^{+5}$ }
             &  $33.7$
			 &  $32.1$
			 &  $59.1$
			 &  $81.2$
\\
\hline
{\small $\Delta_{1600}\rightarrow \pi N_{939} $ }
	 & { $\left(61\pm 26\right)_{-10}^{+26} $ }
             &  $0.116$
			 &  $0.503$
			 &  $74.2$
			 &  $55.7$
\\
\hline
{\small $\Delta_{1620}\rightarrow \pi N_{939} $}
	 & { $\left(38\pm 8 \right)_{-6}^{+8}$ }
             &  $10.4$
			 &  $14.6$
			 &  $4.82$
			 &  $74.8$
\\
\hline
{\small $\Delta_{1700}\rightarrow \pi N_{939} $ }
	 & { $ \left(45\pm 15\right)_{-10}^{+20}$ }
             &  $2.92$
			 &  $3.10$
			 &  $7.12$
			 &  $14.4$
\\
\hline
\end{tabular}
\end{table}

Following the studies of nucleon form factors addressed in the 
previous section, the Graz group has 
recently performed calculations of (pionic) decay
widths of $N$ and $\Delta$ resonances with CQMs in relativistic
point-form quantum mechanics. A first report thereof was given in ref.
\cite{Melde:2002ga}. In table 3 we provide a comparison of some 
results for pionic decay widths of $N$ and $\Delta$ resonances from
the GBE and BCN OGE CQMs. So far one has only employed a decay
operator according to a relativistic generalization of the elementary
emission model (EEM), i.e. a one-body operator.
Relativistic boost effects, however, have been
included exactly in the calculation of matrix elements. By 
comparison to a standard nonrelativistic calculation one has again
found that relativistic effects are of considerable importance, 
with their magnitudes exceeding by far the influences of dynamical
ingredients in the wave 
functions. Still, at this instance, one has not yet arrived at a 
satisfactory explanation of the decay widths in accordance with 
existing experimental data. Here, the problems appear 
much more intricate than anticipated. Due to the involvement of 
resonance wave functions in evaluating the transition matrix elements 
one might tentatively need a more elaborate description of the 
resonance states with finite rather than zero widths. This means that 
one should include the couplings to the decay channels in the 
resonance wave functions. In a microscopic approach, ideally, one 
should provide additional configurations beyond three-quark states 
explicitly or 
at least take care of their roles in an effective description.
Similarly one must consider further refinements in the decay
operator and/or alternative mechanisms for the meson emission.

\section{Conclusions}

From the evidences highlighted above, a number of important
conclusions can be drawn on the performance and adequacy of CQMs (as 
effective models of QCD for low-energy hadronic physics). These 
insights will be important for the future development of the CQM
approach and, more generally, for the physics of low- and
intermediate-energy hadron phenomena.

Regarding spectroscopy a further improvement of CQMs is most
desirable. Certainly, with the GBE CQM (and to a lesser quality also 
with the II CQM) a description of all light and strange baryons is
at hand within a single model\footnote{We recall that, e.g., for
the Capstick-Isgur OGE CQM \protect\cite{CI} two separate 
parametrizations are necessary, one for positive-parity and one for 
negative-parity states. There is not a unique parameter set allowing
for a simultaneous reproduction of all states in a uniform manner.}.
It would be satisfying
to extend the existing models in order to reach an even more
comprehensive unified framework for the treatment of baryons (including
also the heavy flavors) and mesons. From the applications considered 
here, one might be content with the behaviour of ground 
states (as \{$QQQ$\} bound states) but the treatment of resonances
must certainly be made more realistic. In this respect it is also not
clear how far the concept of CQMs can be pushed up in excitation
energy, even under the inclusion of further, say mesonic, degrees
of freedom. The assumption of the constituent quark itself, as a
quasiparticle of low-energy QCD, could soon break down in higher
excitations of hadrons, namely, when chiral symmetry gets restored.

From dealing with hadron reactions in CQMs (in our case 
lepton scattering on nucleons and strong decays of baryon
resonances) it has become fairly evident that a relativistic
formulation has to be employed. In this light, any nonrelativistic
treatment must be considered as inadequate from the very beginning.
Still, it appears viable to work with a finite number of degrees
of freedom instead of a genuine field-theoretic framework. 
This paves the way for approaches following Poincar\'e-invariant 
quantum mechanics. While being well founded axiomatically,
it seems they also provide a proper tool for 
fulfilling Lorentz covariance and allow at the same time the 
application of refined few-body techniques to solve a variety of 
few-quark problems without having to resort to drastic approximations.
Over the past decades few-body physics has been especially successful 
in accurate descriptions of nuclear phenomena. For that purpose a 
nonrelativistic quantum theory was mostly sufficient. In subnuclear 
physics, when dealing with a microscopic description of hadrons as 
few-quark systems, this is no longer the case, and we have to employ 
relativistic approaches. Their solution usually requires much more 
technical efforts, and it represents a challenge to apply them
for all kind of hadron phenomena. We may well be confident
that the corresponding studies will reward us with new exciting
flavours in the traditional field of few-body physics.

\begin{acknowledge}
I want to express my sincere gratitude to Teresa Pe\~na and her team 
for organizing the 2002 Lisbon Workshop and for dedicating it to Peter 
Sauer. Indeed, it is justified to acknowledge Peter's contributions
to few-body physics over the years by a topical workshop. I have been
knowing him since the early 70's and I have always appreciated him as
a critical colleague and an enthusiastic as well as helpful collaborator.
The scientific links between Hanover and Graz have produced a number
of essential results and also several friendly personal relationships.
I feel uncomfortable when thinking about a future where all this could
be lost due to the retirement of Peter and a possible disruption of
the Hanover few-body-physics group.
\newline
The results discussed in this work
have mostly originated from combined efforts of the few-body physics
group in Graz, partly in collaboration with the Pavia group (Sigfrido
Boffi and Marco Radici) and with Bill Klink from the University of
Iowa. It is a pleasure to acknowledge this fruitful collaboration
and here above all the individual contributions by Robert Wagenbrunn,
Katharina Berger, and Thomas Melde. This work was supported by the
Austrian Science Fund, project no. P14806-TPH. 
\end{acknowledge}

\end{document}